\documentstyle[12pt]{article}
\advance \topmargin by -\headheight
\advance \topmargin by -\headsep
\evensidemargin \oddsidemargin
\title{Gravitational Theory without the Cosmological Constant Problem}
\author{E.I.Guendelman\thanks{GUENDEL@BGUmail.BGU.AC.IL} and
        A.B.Kaganovich \thanks{ALEXK@BGUmail.BGU.AC.IL}}
\date {Physics Department, Ben Gurion University of the Negev, 
   Beer  Sheva 84105, Israel} 

\begin{document}
\maketitle
\begin{abstract}
We develop the principle of nongravitating vacuum energy, which is 
implemented by changing the measure of integration from $\sqrt{-g}d^{D}x$ 
to an integration in an internal space of $D$ scalar fields 
$\varphi_{a}$. As a consequence of such a choice  of the measure, the 
matter Lagrangian $L_{m}$ can be changed by adding
 a constant while no cosmological term is induced. Here we develop this 
idea to build a new theory which is formulated through
the first order formalism, i.e. using vielbein $e_{a}^{\mu}$ and spin 
connection $\omega_{\mu}^{ab}$ ($a,b=1,2,...D$) as independent 
variables. The equations obtained from the variation of $e_{a}^{\mu}$ 
and the fields $\varphi_{a}$ imply the existence of a 
nontrivial constraint.  This approach
can be made consistent with invariance under arbitrary diffeomorphisms in 
the internal space of scalar fields $\varphi_{a}$ (as well as in ordinary 
space-time), provided that the matter model is chosen so as to satisfy 
the above mentioned constraint. If the matter model is not chosen so as 
to satisfy automatically this constraint, the diffeomorphism invariance 
in the internal space is broken. In this case the constraint is 
dynamically implemented by the degrees of freedom that become physical 
due to the breaking of the internal diffeomorphism invariance. However,  
this constraint always dictates the vanishing of the cosmological 
constant term and the gravitational equations in the vacuum coincide 
with vacuum Einstein's equations with zero cosmological constant. The 
requirement that the internal diffeomorphisms be a 
symmetry of the theory points towards the unification of forces in nature 
like in the Kaluza-Klein scheme.

\end{abstract}

\pagebreak

\section{Introduction}

\bigskip

In 1917, Einstein realized \cite{Ein}, that his field equations can be 
modified by introducing the "cosmological constant term". This 
"$\Lambda$-term" appears in the Einstein's equations in the form
\begin{equation}
 R_{\mu\nu}-\frac{1}{2}g_{\mu\nu}R-\Lambda g_{\mu\nu}=
 \frac{\kappa}{2}T_{\mu\nu}
        \label{Ein}
\end{equation}
where $\kappa\equiv 16\pi G$. Although Einstein considered the 
introduction of such a term a mistake, the fact is that such term does not 
violate any symmetry. Furthermore, quantum field theory (QFT) predicts the 
existence a vacuum energy due to the zero point fluctuations, which gives 
an infinite contribution to $T_{\mu\nu}$ which is of the form 
$g_{\mu\nu}\rho_{0}$, ($\rho_{0}=const$), that is, indistinguishable from 
the $\Lambda$-term. Even if the infinity problem could be avoided,  
QFT  naturally predicts a very large $\Lambda$-term, since on purely 
dimensional ground, QCD would give a vacuum energy of order $1Gev^{4}$ 
and in quantum gravity one expects $10^{76}Gev^{4}$, while observations 
require the vacuum energy to be less then $10^{-46}Gev^{4}$. For a 
historic overview see \cite{Nov} and for reviews of the modern attempts 
to solve this puzzle see \cite{WNg}.

In this paper we will develop an approach where as a consequence of a 
nontrivial constraint imposed by the variational principle, which has a 
highly geometrical motivation, any $\Lambda$-term is forbidden. When this 
constraint  is satisfied in an automatic form by the matter models, we 
have an additional local symmetry in the model. The triviality of the 
constraint or the associated local gauge symmetry implies the physical 
irrelevance of certain degrees of freedom. It should be pointed out that 
the vanishing of the cosmological term is achieved even if the constraint 
is nontrivially implemented. In this case the degrees of freedom 
mentioned above become nontrivial and are dynamically active in the 
mechanism that eliminates the cosmological constant.

This approach is based on a paper by us \cite{GK}, where the "principle 
of nongravitating vacuum energy (NGVE)" was formulated.
There the usual measure of integration that is $\sqrt{-g}$, was changed 
by another scalar density $\Phi$ which is also a total derivative, built 
from $D$ scalar fields (if D is the dimension of space-time). In an 
explicit form:
\begin{equation}
\Phi \equiv \varepsilon_{a_{1}a_{2}\ldots a_{D}}
\varepsilon^{\alpha_{1}\alpha_{2}\ldots \alpha_{D}}
(\partial_{\alpha_{1}}\varphi_{a_{1}})
(\partial_{\alpha_{2}}\varphi_{a_{2}}) \ldots
(\partial_{\alpha_{D}}\varphi_{a_{D}})  
\label{Fish}
\end{equation}
where $\varphi_{a}, (a=1,2,...D)$ are scalar fields.
In this case $\int L_{m}\Phi d^{D}x$ is invariant under the change 
$L_{m}\longrightarrow L_{m}+constant$, since then we just add to the 
integrand 
 $L_{m}\Phi$ a total derivative term. We should then remember 
that 
usually the cosmological constant piece in (\ref{Ein}) is generated from 
a term of the form $\Lambda\int\sqrt{-g} d^{D}x$, which with the change 
of measure becomes an irrelevant total divergence.
In spite of this, in such a model an integration constant that plays a 
role which resembles that of a cosmological term, appears in the equations
(although for nonvanishing values of this integration constant, maximally 
symmetric spaces are not available\cite{GK}). In addition, the equations 
deviate from those of general relativity and a new physical massless 
"dilaton" appears, with the corresponding phenomenological problems.

In contrast, here we will find that when formulating the theory in a way 
which is invariant under diffeomorphisms in the manifold of fields 
$\varphi_{a}$, then no term that plays the role of a cosmological 
constant term can appear. Together with this, no propagating dilaton 
appears.  This is achieved in a simple 
way by formulating the model in terms of vielbeins and allowing the 
possibility of torsion. This is a quite natural approach since 
vielbeins and torsion appear in any case if fermions are introduced.         

In the context of this formulation of the theory, the local symmetry 
mentioned before, is actually the group of diffeomorphisms in the space 
of the scalar fields $\varphi_{a}$. If this group is indeed a symmetry, 
the vanishing of the cosmological constant is a trivial consequence. If 
this group is not a symmetry, still the variation with respect to these 
degrees of freedom leads to the constraint that makes the cosmological 
constant vanishing.

As was mentioned above, the key idea of the theory is replacing the measure
$\sqrt{-g} d^{D}x$ by $\Phi d^{D}x$, $\Phi$ being given by (\ref{Fish}) and 
we are then led to the following action for gravity plus matter:
\begin{equation}
S =\int L\Phi d^{D}x\equiv \int (-\frac{1}{\kappa}R + 
L_{m})\Phi d^{D}x
        \label{Action}
\end{equation}
where $R$ is the scalar curvature. To define $R$, we can use the standard 
Riemannian definition in terms of $g_{\mu\nu}$. This leads to the theory 
studied in Ref.\cite{GK}.

A different approach, which will be shown in this paper to be {\em 
physically 
inequivalent}, is to allow a more general form for $R$ which allows for 
the possibility of torsion. In this case we define \cite{Tor}
\begin{equation}
R(\omega ,e) =e^{a\mu}e^{b\nu}R_{\mu\nu ab}(\omega), 
\label{Scurv}
\end {equation}
\begin{equation}
R_{\mu\nu ab}(\omega)=\partial_{\mu}\omega_{\nu ab}
-\partial_{\nu}\omega_{\mu ab}+(\omega_{\mu a}^{c}\omega_{\nu cb}
-\omega_{\nu a}^{c}\omega_{\mu cb})
        \label{Curv}
\end{equation}
where $e^{a\mu}=\eta^{ab}e_{b}^{\mu}$, $\eta^{ab}$ is the diagonal 
$D\times D$ 
matrix with elements $+1, -1,...-1$ on the diagonal, $e_{a}^{\mu}$ 
are the vielbeins and $\omega_{\mu}^{ab}$ ($a,b=1,2,...D$) is 
the spin connection. The matter Lagrangian $L_{m}$ that appears in 
eq.(\ref{Action}) still does not depend on the  scalar fields $\varphi_{a}$ 
and it is now function of matter fields, vielbeins and spin 
connection, considered as independent fields. We assume for simplicity 
that $L_{m}$ does not depend on the derivatives of vielbeins and spin
connection.

As it is well known\cite{W}, if fermions contribute to $L_{m}$, the 
vielbein formalism becomes unavoidable anyway. This could be regarded as 
an argument to view the use of (\ref{Scurv}) and (\ref{Curv}) as a more 
fundamental starting 
point than that of using the Riemannian definition for $R$. As we will 
see, in the NGVE theory studied here $R(\omega ,e)\neq$ Riemann scalar, 
even in the case $L_{m}=0$.

\pagebreak

\section{General features of the NGVE-theory}

\bigskip
In this section we study the general features of the NGVE-theory, which 
are consequences only of the fact that the scalar fields $\varphi_{a}$ enter 
just in the measure of integration and not in the total Lagrangian density
$L\equiv - \frac{1}{\kappa}R +L_{m}$.
First notice that $\Phi$ is the Jacobian of the mapping
$\varphi_{a}=\varphi_{a}(x^{\alpha})$, $a=1,2,\ldots,D$. If
this mapping is nonsingular $(\Phi\neq 0)$ then (at least locally) there
is  the inverse mapping $x^{\alpha}=x^{\alpha}(\varphi_{a})$, $\alpha=
0,1\ldots,D-1$.  Since $\Phi d^{D}x = D!
d\varphi_{1}\wedge d\varphi_{2}\wedge\ldots\wedge d\varphi_{D}$  we can
think  $\Phi d^{D}x$ as integrating in the internal space
variables $\varphi_{a}$.  Besides, if $\Phi \neq 0$ then there is a
coordinate frame where the coordinates are the scalar fields themselves.

The field $\Phi$ is invariant under the volume preserving diffeomorphisms
in internal space: $\varphi^{\prime}_{a}=\varphi^{\prime}_{a}(\varphi_{b})$
where \begin{equation}
\varepsilon_{a_{1}a_{2}\ldots
a_{D}}\frac{\partial{\varphi^{\prime}}_{b_{1}}}{\partial{\varphi}_{a_{1}}}
\frac{\partial{\varphi^{\prime}}_{b_{2}}}{\partial{\varphi}_{a_{2}}}\ldots
\frac{\partial{\varphi^{\prime}}_{b_{D}}}{\partial{\varphi}_{a_{D}}}=
 \varepsilon_{b_{1}b_{2}\ldots b_{D}}
        \label{epsilon}
\end{equation}
 
Such infinite dimensional symmetry leads to an infinite number of 
conservation laws.
To see this, notice that from the volume preserving
symmetries $\varphi^{\prime}_{a}=\varphi^{\prime}_{a}(\varphi_{b})$
defined by eq.(\ref{epsilon}) which for the infinitesimal case implies
\begin{equation}
        \varphi^{\prime}_{a}=\varphi_{a}+
        \lambda\varepsilon^{aa_{1}\ldots a_{D}}
        \frac{\partial F_{a_{1}a_{2}\ldots a_{D-1}}(\varphi_{b})}
        {\partial\varphi_{a_{D}}}
        \label{infphi}
\end{equation}

($\lambda\ll 1$), we obtain, through Noether's theorem the following
conserved quantities
\begin{equation}
        j_{V}^{\mu}=A_{a}^{\mu}(-\frac{1}{\kappa}R+L)
        \varepsilon_{aa_{1}\ldots a_{D}}
        \frac{\partial F_{a_{1}a_{2}\ldots a_{D-1}}(\varphi_{b})}
        {\partial\varphi_{a_{D}}}
        \label{jv}
\end{equation}

We now want to notice that the form of the action (\ref{Action}) implies the 
existence of a very 
special set of equations. These are the equations of motion obtained by 
variation of the 
action (\ref{Action}) with respect to the scalar fields $\varphi_{b}$ and 
they are
 \begin{equation}
 A_{b}^{\mu}\partial_{\mu}(-\frac{1}{\kappa}R + L_{m} ) = 0
 \label{Inv}
 \end{equation}
  where
  \begin{equation}
  A_{b}^{\mu}\equiv \varepsilon_{a_{1}a_{2}\ldots a_{D-1}b}
\varepsilon^{\alpha_{1}\alpha_{2}\ldots \alpha_{D-1}\mu}
(\partial_{\alpha_{1}}\varphi_{a_{1}})
(\partial_{\alpha_{2}}\varphi_{a_{2}}) \ldots
(\partial_{\alpha_{D-1}}\varphi_{a_{D-1}})
        \label{A}
  \end{equation}
It follows from (\ref{Fish}) that
$A_{b}^{\mu}\partial_{\mu}\varphi_{b^{\prime}}=D^{-1}\delta_{bb^{\prime}}\Phi$
and taking the determinant of
both sides, we get $det (A_{b}^{\mu}) = \frac{D^{-D}}{D!}\Phi^{D-1}$.
Therefore if $\Phi\neq 0$, which we will assume in what follows, the
only solution for (\ref{Inv}) is
\begin{equation}
L\equiv- \frac{1}{\kappa} R + L_{m} = constant\equiv M \label{IntM}
\end{equation}

Finally, we show that the same structure of this action, which leads 
to  the very special set of equations displayed above, is associated 
with another, even more puzzling set of symmetries than the volume 
preserving diffeomorphisms. In fact, let us consider the following 
infinitesimal shift of the fields
$\varphi_{a}$ by an arbitrary infinitesimal function of the total Lagrangian
density
$L\equiv-\frac{1}{\kappa}R+L_{m}$, that is
\begin{equation}
        \varphi^{\prime}_{a}=\varphi_{a}+\epsilon g_{a}(L),   \epsilon\ll 1
        \label{L}
        \end{equation}
In this case the action is transformed according to
\begin{equation}
        \delta S=\epsilon D\int A_{a}^{\mu}L\partial_{\mu}
        g_{a}(L)d^{D}x=\epsilon\int\partial_{\mu}\Omega^{\mu}d^{D}x
        \label{deltaS}
\end{equation}
where $\Omega^{\mu}\equiv DA_{a}^{\mu}f_{a}(L)$ and $f_{a}(L)$ being defined
from $g_{a}(L)$ through the equation $L\frac{dg_{a}}{dL}=\frac{df_{a}}{dL}$.
To obtain the last expression in the equation(\ref{deltaS}) it is
necessary to note that $\partial_{\mu}A_{a}^{\mu}\equiv 0$. By means of
the Noether's theorem, this symmetry leads to the conserved current
\begin{equation}
        j_{L}^{\mu}=A_{a}^{\mu}(Lg_{a}-f_{a})\equiv
        A_{a}^{\mu}\int_{L_{0}}^{L}g_{a}(L^{\prime})dL^{\prime}
        \label{jl}
\end{equation}
The existence of this symmetry depends crucially on the independence of 
the Lagrangian density L on the scalar fields that define the measure.
In fact, the existence of this symmetry could be used to justify the 
expectation that quantum corrections would keep that basic structure, 
provided it is present at the tree level. 
 \pagebreak

\section{The NGVE-theory - Riemannian approach}

\bigskip

Before studying the case when the definitions (\ref{Scurv}) and  
(\ref{Curv}) are used we 
will review the model studied in \cite{GK}, where R in the action 
(\ref{Action}) is the Riemannian one and $L_{m}=L_{m}(g_{\mu\nu}, matter 
fields)$.

Variation of $S_{g}\equiv-\frac{1}{\kappa}\int R\Phi d^{D}x$
with respect to
$g^{\mu\nu}$ leads to the result
\begin{equation}
        \delta S_{g}=-\frac{1}{\kappa}\int \Phi[R_{\mu\nu}+
        (g_{\mu\nu}\Box-\nabla_{\mu}\nabla_{\nu})]\delta g^{\mu\nu}d^{D}x
        \label{Andro}
\end{equation}

In order to perform the correct integration by parts we have to make
use
the scalar field $\chi\equiv\frac{\Phi}{\sqrt{-g}}$, which is invariant
under continuous general coordinate transformations, instead of the
scalar density
$\Phi$.  Then integrating by parts and ignoring a total derivative term
which has the form $\partial_{\alpha}(\sqrt{-g}P^{\alpha})$, where
$P^{\alpha}$ is a vector field, we get
\begin{equation}
        \frac{\delta S_{g}}{\delta 
g^{\mu\nu}}=-\frac{1}{\kappa}\sqrt{-g}[\chi
        R_{\mu\nu}+g_{\mu\nu}\Box\chi-\chi_{,\mu;\nu}]
        \label{var}
\end{equation}

In a similar way varying the matter part of the action (\ref{Action})
with
respect to $g^{\mu\nu}$ and making use the scalar field $\chi$ we can
express a result in terms of the standard matter energy-momentum tensor
$T_{\mu\nu}\equiv\frac{2}{\sqrt{-g}}\frac{\partial(\sqrt{-g}L_{m})}{\partial
g^{\mu\nu}}$. Then after some algebraic manipulations we get instead of
Einstein's equations
\begin{equation}
        G_{\mu\nu}=\frac{\kappa}{2} [T_{\mu\nu} -
        \frac{1}{2}g_{\mu\nu}(T^{\alpha}_{\alpha}+(D-2)L_{m})]+
        \frac{1}{\chi}(\frac{D-3}{2}g_{\mu\nu}\Box\chi +\chi ,_{\mu ;\nu})
                \label{Ein1}
\end{equation}
where $G_{\mu\nu}\equiv R_{\mu\nu}-\frac{1}{2}Rg_{\mu\nu}$.

By contracting (\ref{Ein1}) and using (\ref{IntM}), we get
\begin{equation}
\Box\chi-\frac{\kappa}{D-1}[M+\frac{1}{2}(T^{\alpha}_{\alpha}
+(D-2)L_{m})]\chi=0
\label{Scal}
\end{equation}

By using eq.(\ref{Scal}) we can now exclude
$T^{\alpha}_{\alpha}+(D-2)L_{m}$ from eq.(\ref{Ein1}):
\begin{equation}
G_{\mu\nu}=\frac{\kappa}{2} [T_{\mu\nu}+Mg_{\mu\nu}]+
        \frac{1}{\chi}[\chi_{,\mu;\nu}-g_{\mu\nu}\Box\chi]
        \label{M-eq}
\end{equation}

Notice that eqs.(\ref{Inv}) and (\ref{Ein1}) are invariant under the
addition
to $L_{m}$ a constant piece, since the combination $T_{\mu\nu}-
\frac{1}{2}g_{\mu\nu}[T^{\alpha}_{\alpha}+(D-2)L_{m}]$ is invariant.

It is very important to note that the terms depending on the matter
fields
in eq.(\ref{M-eq}) as well as in eq.(\ref{Ein1}) do not contain $\chi$
-field, in contrast to the usual scalar-tensor theories, like
Brans-Dicke
theory.  As a result of this feature of the NGVE- theory,
the
gravitational constant does not suffer space-time variations.
However, the matter
energy-momentum tensor $T_{\mu\nu}$ is not conserved.  Actually,
 taking the
covariant divergence of both sides of eq.(\ref{M-eq})
 and using the identity
$\chi^{,\alpha} _{;\nu;\alpha}=(\Box\chi),_{\nu}
+\chi^{,\alpha}R_{\alpha\nu}$, eqs.(\ref{M-eq}) and
(\ref{Scal}),
we get the equation of matter non conservation
\begin{equation}
T_{\mu\nu}\:^{;\mu}=-2\frac{\partial L_{m}}{\partial g^{\mu\nu}}
g^{\mu\alpha}\partial_{\alpha}ln\chi
\label{Noncons}
\end{equation}

We are interested now in studying the question whether there is an
Einstein sector of solutions, that is are there solutions that satisfy
Einstein's equations? First of all we see that eqs.(\ref{M-eq}) coincide with
Einstein's equations  only if the
$\chi$-field is a constant.  From eq.(\ref{Scal}) we conclude that this is
possible
only if an essential restriction on the matter model is imposed
\begin{equation}
2M +T^{\alpha}_{\alpha}+(D-2)L_{m}\equiv 2[g^{\mu\nu}\frac{\partial
(L_{m}-M)}{\partial g^{\mu\nu}}-(L_{m}-M)]=0,
\label{Sector}
\end{equation}
which means that $L_{m}-M$ is an homogeneous function of $g^{\mu\nu}$ of
degree one, in any dimension.  If condition (\ref{Sector}) is satisfied
then the equations of motion allow solutions of GR to be solutions of
the
model, that is $\chi = constant$ and $G_{\mu\nu}=
\frac{\kappa}{2}T_{\mu\nu}+Mg_{\mu\nu}$.
It is interesting to observe that when condition (\ref{Sector}) is
satisfied, a new symmetry of the action(\ref{Action}) appears. We will call
this symmetry `'Einstein
symmetry`` (because (\ref{Sector}) leads to the existence of an Einstein
sector of solutions).
Such symmetry consists of the scalings
\begin{equation}
        g^{\mu\nu}\rightarrow \lambda g^{\mu\nu}
        \label{eg}
\end{equation}
\begin{equation}
        \varphi_{a}\rightarrow \lambda^{-\frac{1}{D}}\varphi_{a}
        \label{Ephi}
\end{equation}
where $\lambda =const$.
To see that this is indeed a symmetry, note that from definition of scalar
curvature it follows that $R\rightarrow \lambda
R$ when the transformations (\ref{eg}),(\ref{Ephi}) are performed. Since 
condition (\ref{Sector}) means that $L_{m}$ is a homogeneous function of
$g^{\mu\nu}$ of degree 1, we see that under the transformations
(\ref{eg}),,(\ref{Ephi}) the matter Lagrangian $L_{m}\rightarrow \lambda 
L_{m}$. From
this we conclude that (\ref{eg}),(\ref{Ephi}) is indeed a symmetry of the
action(\ref{Action}) when (\ref{Sector}) is satisfied.

The situation described above can be realized for special kinds
of bosonic matter models:

1.Scalar fields without potentials, including fields subjected to non
linear constraints, like the $\sigma$ model.  The general coordinate
invariant action for these cases has the form $S_{m} =\int L_{m}\Phi
d^{D}x$ where $L_{m}=
\frac{1}{2}\sigma,_{\mu}\sigma,_{\nu}g^{\mu\nu}$.

2.Matter consisting of fundamental bosonic strings.  The condition
(\ref{Sector}) can be verified by representing the string action in the
$D$-dimensional form where $g_{\mu\nu}$ plays the role of a background
metric.  For example, bosonic strings, according to our formulation,
where
the measure of integration in a $D$ dimensional space-time is chosen to
be $\Phi d^{D}x$, will be governed by an action of the form:
\begin{equation}
S_{m} =\int
L_{string}\Phi d^Dx,
L_{string}= -T\int d\sigma
d\tau\frac{\delta^{D}(x-X(\sigma,\tau))}{\sqrt{-g}}
\sqrt{det(g_{\mu\nu}X^{\mu}_{,a}X^{\nu}_{,b})}
        \label{String}
\end{equation}
where $\int L_{string}\sqrt{-g}d^{D}x$ would be the action of a string
embedded
in a $D$-dimensional space-time in the standard theory; $a,b$ label
coordinates in the string world sheet and $T$ is the string tension.
Notice that under a scaling (\ref{eg})  (which means that $g_{\mu\nu}
\rightarrow\lambda^{-1}g_{\mu\nu}$),
$L_{string}\rightarrow\lambda^{(D-2)/2}L_{string}$ , therefore
concluding that $L_{string}$ is a homogeneous function of $g^{\mu\nu}$
of degree one, that is eq.(\ref{Sector}) is satisfied only if $D=4$.

3.It is possible to formulate {\em the point particle model\/} of matter
in a way
such that eq.(\ref{Sector}) is satisfied.  This is because for the
free falling point
particle a variety of actions are possible (and are equivalent in the
context of general relativity).  The usual actions are taken to be
$S=-m\int F(y)ds$, where
$y=g_{\alpha\beta}\frac{dX^{\alpha}}{ds}\frac{dX^\beta}{ds}$ and $s$ is
determined to be an affine parameter except if $F=\sqrt{y}$, which is
the
case of reparametrization invariance.  In our model we must take
$S_{m}=-m\int L_{part}\Phi d^{4}x$ with $L_{part}=-m\int
ds\frac{\delta^{4}(x-X(s))}{\sqrt{-g}}F(y(X(s)))$ where $\int
L_{part}\sqrt{-g}d^{4}x$ would be the action of a point particle in 4
dimensions in the usual theory.  For the choice $F=y$, condition
(\ref{Sector}) is satisfied.  Unlike the case of general relativity,
different choices of $F$ lead to unequivalent theories.
Notice that in the case of point particles (taking $F=y$), a geodesic
equation (and therefore the equivalence principle) is satisfied in
terms of
the metric $g^{eff}_{\alpha\beta}\equiv\chi g_{\alpha\beta}$ even if $\chi$
is
not constant. It is interesting also that in the 4-dimensional case
 $g^{eff}_{\alpha\beta}$ is
invariant under the Einstein symmetry described by eqs.(\ref{eg}) and
(\ref{Ephi}).

Notice that the theory as formulated in this section makes sense even 
without condition (\ref{Sector}) being satisfied. In contrast, we will 
see in the next section that when allowing torsion, a condition which 
generalizes the condition (\ref{Sector}), {\em is required} for  the 
consistency of the equations of motion of the theory. 

\pagebreak

\section{The NGVE-theory- Vielbein-Spin Connection Approach}

\bigskip
\subsection{General Consideration}

We are now going to study the theory defined by the action 
(\ref{Action}) in the case that the scalar curvature is defined by 
(\ref{Scurv}),(\ref{Curv}), which means that $R$ may not coincide with the 
Riemannian scalar curvature and as a consequence we do not expect the 
NGVE-Vielbein-Spin Connection (VSC) approach to coincide with the Riemannian 
approach of section 3.

As in section 2, variation with respect to the scalar fields $\varphi_{a}$ 
leads to the equations

\begin{equation}
 A_{a}^{\mu}\partial_{\mu}(-\frac{1}{\kappa}R(e,\omega ) + L_{m}(e,\omega 
, matter fields)) = 0
 \label{V}
 \end{equation}
which implies, if $\Phi \neq 0$, that

\begin{equation}
 -\frac{1}{\kappa}R(e,\omega ) + L_{m}(e,\omega, matter fields) = M
 \label{M}
 \end{equation}

On the other hand, considering the equations obtained from the variation 
of the vielbeins, we get if  $\Phi \neq 0$
\begin{equation}
 -\frac{2}{\kappa}R_{a\mu}(e,\omega)
+\frac{\partial L_{m}}{\partial e^{a\mu}} = 0,
 \label{ET}
 \end{equation}
where
\begin{equation}
 R_{\mu a}(e,\omega)\equiv 
e^{b\nu}R_{\mu\nu ab}(\omega).
\label{RT}
\end{equation}

Notice that eq.(\ref{ET}) is indeed invariant under the shift 
$L_{m}\rightarrow L_{m}+const$.

Since $R(e,\omega)\equiv e^{a\mu}R_{\mu a}(e,\omega)$, we can eliminate 
$R(\omega)$ from the equations (\ref{M}) and (\ref{ET}) after contracting 
the last one with $e_{a\mu}$. As a result we obtain {\em the nontrivial 
constraint}  

\begin{equation}
e^{a\mu}\frac{\partial(L_{m}-M)}{\partial e^{a\mu}}-2(L_{m}-M)=0
\label{C}
 \end{equation}

In the case $L_{m}=L_{m}(g_{\mu,\nu}, matter fields)$ we see that the form 
of the constraint (\ref{C}) coincides  with the condition (\ref{Sector}) 
which 
provides the existence of the Einstein sector of solutions in the Riemannian 
approach. In contrast {\em here it is not a choice but it is a 
consequence of the variational principle.}

The constraint (\ref{C}) has to be satisfied for all components (in the 
functional space) of the function $L_{m}$. In particular, for the 
constant part denoted $<L_{m}>$ we obtain:

\begin{equation}
<L_{m}>-M=0
\label{C1}
 \end{equation}
 Therefore, one of 
the consequences of the  constraint (\ref{C}) is that it {\em dictates 
that the constant part of the matter Lagrangian $<L_{m}>$ is compensated by 
the integration constant $M$.}

We will see that constraint (\ref{C}) can be satisfied in three possible 
ways: (1) automatically, that is from the definition of $L_{m}$, without 
any dynamical consideration; (2) automatically after matter field 
equations {\em only} are used; (3) after all equations are used. All 
three matter model examples of the section 3 belong to case (1). 

As we saw in section 3, in the context of the Riemannian approach, the 
condition (\ref{Sector}) is related to the Einstein symmetry (\ref{eg}), 
(\ref{Ephi}). It is very interesting to see what kind of symmetry of the 
action (\ref{Action}) is associated with the constraint (\ref{C}) in the 
context of the 
VSC- approach. It turns 
out that when the constraint 
(\ref{C}) is satisfied automatically (without using the equations of 
motion of matter) we obtain that a local version of Einstein symmetry holds. 
Furthermore, 
this {\em local Einstein symmetry\/} is nothing but diffeomorphism 
invariance in the space of 
the scalar fields $\varphi_{a}$, which has to be accompanied with a 
conformal transformation of the vielbeins:
\begin{equation}
\varphi_{a}\rightarrow\varphi '_{a}=\varphi '_{a}(\varphi_{b}),
\label{LAS1}
\end{equation}
\begin{equation}
e_{a\mu}\rightarrow e'_{a\mu}=J^{1/2}e_{a\mu}, 
\label{LAS2}
\end{equation}
\begin{equation}
J\equiv Det(\frac{\partial\varphi'_{a}}{\partial\varphi_{b}}).
\label{LAS3}
 \end{equation}

In terms of $g^{\mu\nu}$ and $\Phi$ (and 
$\chi\equiv\frac{\Phi}{\sqrt{-g}}$) this symmetry has the form:
 \begin{equation}
g^{\mu\nu}\rightarrow g'^{\mu\nu}=J^{-1}g^{\mu\nu}
\label{LSI1}
\end{equation}
\begin{equation}
\Phi\rightarrow \Phi ' =J\Phi
\label{LSI2}
\end{equation}
\begin{equation}
\chi\rightarrow \chi ' =J^{1-D/2}\chi
\label{LSI3}
\end{equation}

Since when $\Phi\neq 0$, we have that the transformation 
$\varphi_{a}=\varphi_{a}(x^{\mu})$ is one to one, we obtain that by means 
of (\ref{LSI2}),  $\Phi$ can be transformed to whatever we want, in 
particular $\Phi =\sqrt{-g}$ or, what is the same, $\chi =1$ is a possible 
"gauge" if $\Phi\neq 0$.

\bigskip
\subsection{Torsion in the absence of fermions}

Let us now analyse what is the dependence of $\omega_{\mu}^{ ab}$ on 
$e_{a\mu}$ and $\chi$. As a first step, let us consider the case when 
$L_{m}=L_{m}(g_{\mu\nu}$, matter fields) and the dimensionality of the 
space-time D=4. This of course excludes 
the possibility of fermions, but those can be incorporated without 
qualitative changes in the discussion.

Then the variation of the action (\ref{Action}) with respect to 
$\omega_{\mu}^{ ab}$ gives:
\begin{equation}
\varepsilon^{\mu\nu\lambda\rho}\varepsilon_{abcd}[\chi 
e^{c}_{\lambda}D_{\nu}e^{d}_{\rho}
+\frac{1}{2}e^{c}_{\lambda}e^{d}_{\rho}\chi,_{\nu}]=0, 
\label{T}
 \end{equation}  
where $D_{\nu}e_{a\rho}\equiv\partial_{\nu}e_{a\rho}
+\omega_{\nu a}^{d}e_{d\rho}$

The solution of eq.(\ref{T}) is
\begin{equation}
\omega_{\mu}^{ab}=\omega_{\mu}^{ab}(e)+K_{\mu}^{ab}
\label{RK}
 \end{equation}
where $\omega_{\mu}^{ab}(e)$ is the Riemannian spin 
connection\cite{Tor},\cite{W} and $K_{\mu}^{ab}$ is the contorsion 
tensor\cite{Tor},\cite{W} which in our case is given by
\begin{equation}
K_{\mu}^{ab}=\frac{1}{2}\sigma ,_{\alpha}(e^{a}_{\mu}e^{b\alpha}-
e^{b}_{\mu}e^{a\alpha}),
\label{K}
 \end{equation}
where $\sigma\equiv\ln\chi$. Notice that deviation of the new measure 
$\Phi$ from the GR measure $\sqrt{-g}$ (that is $\chi\neq constant$) is 
the origin of torsion. 

If we insert this into the expression of $\Phi R(\omega ,e)$, we obtain 
\begin{equation}
\Phi R(\omega ,e)\equiv\sqrt{-g}\chi R(\omega ,e)=\sqrt{-g}[\chi 
R(g_{\mu\nu})-6\chi^{1/2}\Box\chi^{1/2}],
 \label{Conf}
 \end{equation} 
where $R(g_{\mu\nu})$ is the Riemannian scalar curvature. The 
conformal coupling form of the scalar field $\chi^{1/2}$ is apparent.
This is not a surprise since the left hand side is invariant under the 
local conformal rescalings (\ref{LAS1})-(\ref{LAS3}) and the conformal 
coupling 
form in the right hand side is the unique conformally invariant coupling 
between a scalar field and the Riemannian scalar curvature. The 
right hand side represents the resulting second order formalism, that is, 
what is obtained  after solving the spin connection in 
terms of the other fields and then replacing the result into the action. 
The appearance of the additional $-6\chi^{1/2}\Box\chi^{1/2}$ term  in 
(\ref{Conf}), which is absent in the approach developed in  
section 3, clearly shows the inequivalence of the two approaches in all 
cases, even when no assumptions are made concerning the validity of the 
local Einstein symmetry (\ref{LAS1})-(\ref{LAS3}) or, what is the same, 
(\ref{LSI1})-(\ref{LSI3}) .
 
This can be seen also by examining the shape of the equations of 
motion, even when the symmetry (\ref{LSI1})-(\ref{LSI3}) is  not 
assumed to hold.	 From 
equations (\ref{ET}), (\ref{RT}), (\ref{Scurv}), (\ref{Curv}), (\ref{RK}) 
and (\ref{K}), we get
\begin{equation}
G_{\mu\nu}(g)+H_{\mu\nu}=\frac{\kappa}{2}(T_{\mu\nu}+Mg_{\mu\nu}),
\label{Eg}
 \end{equation}
where
\begin{equation}
H_{\mu\nu}\equiv 
2\chi^{-1/2}[g_{\mu\nu}\Box\chi^{1/2}-(\chi^{1/2}),_{\mu 
;\nu}]+ \chi^{-1}[4(\chi^{1/2}),_{\mu}(\chi^{1/2}),_{\nu}
-g_{\mu\nu}(\chi^{1/2}),_{\alpha}(\chi^{1/2})^{,\alpha}]
 \label{H}
 \end{equation}
and $G_{\mu\nu}(g)\equiv R_{\mu\nu}(g)-\frac{1}{2}g_{\mu\nu}R(g)$ with 
$R_{\mu\nu}(g)$ and $R(g)$ being the Riemannian Ricci tensor and scalar. 
Taking the trace of (\ref{Eg}), we get
\begin{equation}
\Box\chi^{1/2}-\frac{1}{6}[R(g)+
\frac{\kappa}{2}(T^{\alpha}_{\alpha}+4M)]\chi^{1/2}=0
 \label{ST}
 \end{equation}
Using that
 \begin{equation}
 T^{\alpha}_{\alpha}=e^{a\mu}\frac{\partial L_{m}}{\partial  
e^{a\mu}}-4L_{m}
\label{tr}
 \end{equation}
and constraint (\ref{C}), we get
 \begin{equation}
\Box\chi^{1/2}-\frac{1}{6}[R(g)-\kappa (L_{m}-M)]\chi^{1/2}=0
 \label{SL}
 \end{equation}

As we expected $\chi^{1/2}$ has an equation which is of the conformally 
coupled type. 

In the vacuum (that is taking into account only constant 
part of $L_{m}$), due to the constraint (\ref{C1}), eq.(\ref{SL}) takes 
the form 
\begin{equation}
\Box\chi^{1/2}-\frac{1}{6}R(g)\chi^{1/2}=0
 \label{SS1}
 \end{equation}
We can see then that eqs.(\ref{SS1}) and (\ref{Eg}) are invariant under 
the transformations (\ref{LSI1})-(\ref{LSI3}) (which in such a case play 
the role of conformal transformations). Therefore, $\chi$-field can be 
transformed into a constant and the resulting equations (\ref{Eg}) 
become just vacuum Einstein's equations with zero cosmological constant.     
As an example how this is realized in a concrete model, see section 4.4. 

From eqs.(\ref{Eg}), (\ref{H}) and (\ref{ET}) we get the equation of 
matter nonconservation 
\begin{equation}
T_{\mu\nu}\:^{;\mu}=-\frac{\partial L_{m}}{\partial e_{a\mu}}
g_{\mu\nu}e_{a}^{\alpha}\partial_{\alpha}ln\chi,
\label{Nonconse}
\end{equation}
where semicolon means covariant derivative in the Riemannian space-time 
with a metric $g_{\mu\nu}$. This equation coincides with 
eq.(\ref{Noncons}) in 
the case where $L_{m}$ depends on vielbeins only through $g_{\mu\nu}$. 
In cases where the local Einstein symmetry (\ref{LAS1})-(\ref{LAS3}) 
(or, what is the same, (\ref{LSI1})-(\ref{LSI3})) holds,  
the $\chi$-field
can be transformed into a constant and then eqs.(\ref{Nonconse}) becomes 
equations of covariant conservation of the energy-momentum tensor.
 \bigskip

\subsection{Study of constraint in fermionic models}
As it is well known \cite{Tor},\cite{W}, one of the most attractive 
features of the vielbein formalism is its ability to incorporate fermions 
in the context of generally coordinate invariant theories.

 The simplest example of a fermion is that of spin $1/2$ particles. In 
this case we regard the spinor field $\Psi$ as a general coordinate 
scalar and transforming nontrivially with respect to local Lorentz 
transformation according to the spin $1/2$ representation of the Lorentz 
group.

Considering the hermitian action (which allows for the possibility of 
fermion self interactions) of the form
\begin{equation}
S_{f}=\int L_{f}\Phi d^{4}x
 \label{AF}
 \end{equation}
where
\begin{equation}
 L_{f}=\frac{i}{2}\overline{\Psi}[\gamma^{a}e_{a}^{\mu}(\overrightarrow{\partial}
_{\mu}+\frac{1}{2}\omega_{\mu}^{cd}\sigma_{cd})
-(\overleftarrow{\partial}_{\mu}+\frac{1}{2}\omega_{\mu}^{cd}\sigma_{cd})
\gamma^{a}e_{a}^{\mu}]\Psi+U(\overline{\Psi}\Psi) 
 \label{LF}
 \end{equation}
Here $\sigma_{cd}\equiv \frac{1}{4}[\gamma_{c},\gamma_{d}]$.

Again, $\omega_{\mu}^{cd}$ should be determined by the equation obtained 
from the variation of the full action with respect to $\omega_{\mu}^{cd}$. 
This in general will give rise to additional contribution to the torsion, 
as it is well known\cite{Tor},\cite{W}.

Here, in the context of the matter model (\ref{AF}),(\ref{LF}) we focus on 
the conditions where  the constraint (\ref{C}) is satisfied, while $\chi$ 
remains unspecified (i.e. remains unphysical).

From (\ref{LF}) and using the equations of motion derived from the 
action (\ref{AF}),(\ref{LF}), we get
\begin{equation}
e_{a}^{\mu}\frac{\partial L_{f}}{\partial e_{a}^{\mu}}-2L_{f}=
\overline{\Psi}\Psi U'-2U,
 \label{CF}
 \end{equation} 
where $U'$ is the derivative of $U$ with respect to its argument. We see 
that the constraint (\ref{C}) is satisfied on the mass shell (since the 
fermion equations of motion are used) with $M=0$ for $L_{f}$ 
defined by eq.(\ref{LF}) if, for example, $U=c(\overline{\Psi}\Psi)^{2}$. 
Any other quartic interaction, like 
$\overline{\Psi}\gamma_{a}\Psi\overline{\Psi}\gamma^{a}\Psi$,  
$\overline{\Psi}\sigma_{ab}\Psi\overline{\Psi}\sigma^{ab}\Psi$, 
$(\overline{\Psi}\gamma_{5}\Psi)^{2}$, etc. would also satisfy the 
constraint (\ref{C}) on the mass shell with $M=0$. In particular, the 
Nambu - Jona-Lasinio 
model\cite{NJL} would also satisfy the constraint (\ref{C}) on the mass 
shell with $M=0$. 

It is interesting to compare these kind of fermionic models where the 
constraint (\ref{C}) is satisfied with $M=0$, with the models discussed 
already at the end of the section 3. Here the constraint is satisfied 
only on the mass shell while in those previous examples the
constraint (\ref{C}) is satisfied automatically, using only the 
definition of the Lagrangian.

\bigskip
\subsection{Example with scalar field}

\bigskip
Now let us consider cases when the constraint (\ref{C}) is not satisfied
without restrictions on the dynamics of the matter fields. Nevertheless, 
the constraint (\ref{C}) holds as a consequence of the variational 
principle in any situation.

A simple case where the constraint (\ref{C}) is not automatic is the case 
of a scalar field with a nontrivial potential $V(\phi)$. In this case the 
constraint (\ref{C}) implies
\begin{equation}
V(\phi)+M=0
 \label{CS}
 \end{equation}
Therefore we conclude that, provided $\Phi\neq 0$, there is no dynamics for 
the theory of a single scalar field, since constraint (\ref{CS}) forces 
this scalar field to be a constant. This means that the effective 
cosmological constant $V(\phi)+M$ in the equations(\ref{Eg}) vanishes 
identically provided $\Phi\neq 0$.

 The constraint (\ref{CS}) has to be 
solved together with the equation of motion
\begin{equation}
\Box\phi+\sigma ,_{\mu}\phi ^{,\mu}+\frac{\partial V}{\partial\phi}=0,
 \label{SE}
 \end{equation} 
where $\sigma =\ln\chi$. From eqs.(\ref{CS}) and (\ref{SE}) we conclude 
that the $\phi$ -field has to be 
located at an extremum of the potential 
$V(\phi)$. Since the constraint(\ref{CS}) eliminates the dynamics of the 
scalar field $\phi$, we cannot really say that we have a situation where 
the symmetry (\ref{LAS1})-(\ref{LAS3}) (or, what is the same, in the 
form (\ref{LSI1})-(\ref{LSI3}) is actually broken, since after solving 
the constraint together with the equation of motion (i.e. on the mass 
shell) the symmetry remains true.

 Then using the constraint 
(\ref{CS}) in the equation of motion for  $\chi^{1/2}$ (\ref{SL}), we get
\begin{equation}
\Box\chi^{1/2}-\frac{1}{6}R(g)\chi^{1/2}=0
 \label{SS}
 \end{equation}
By using the obvious conformal invariance of eq.(\ref{SS})  and of all 
other equations, the $\chi$-field 
can be transformed into a constant, for example $1$ (the correspondent 
conformal transformation  is in fact the particular case of the local 
Einstein symmetry (\ref{LSI1}),(\ref{LSI2}),(\ref{LSI3}) with 
$J(\varphi_{a}(x))=\chi(x)$). Notice that in this simple matter model 
eq.(\ref{Nonconse}) takes the trivial form 0=0.

\pagebreak

\section{The incorporation of Vector Bosons into the NGVE-theory in the 
VSC-approach}

\bigskip
\subsection{General notions}

\bigskip
As it is well known, interactions between elementary particles appear to 
be well described by the exchange of  vector bosons. The incorporation of 
vector bosons is therefore an important subject which has to be dealt with 
in the context of the new gravitational theory developed in this paper.

As we have seen in the case of the point particle models, different 
formulations of a matter model which in the case of GR are physically 
equivalent, can in fact be the origin of inequivalent theories when 
formulated in the framework of the NGVE-theory. As we will see in this 
chapter, a similar situation arises in the case of vector bosons. We will 
discuss here (and in the next section) several options, some consistent with 
local Einstein symmetry and others which are not. 
As it is well known, the vielbein formalism allows us to regard a vector 
in different ways: 
(i) GVLS: a vector under general coordinate transformations, while being 
a scalar under local Lorentz transformations.
(ii) GSLV: a scalar under general coordinate transformations, while 
being a vector under local Lorentz transformations.
Let $A_{\mu}$ be a GVLS. We can then always define a GSLV as 
$A_{a}=e_{a}^{\mu}A_{\mu}$.

\bigskip
\subsection{Model of vector boson with the local Einstein symmetry}

\bigskip
Here we will choose the GSLV variables as the fundamental Lagrangian 
variables.
Defining the Lorentz tensor and generally coordinate invariant scalar field 
strength
 \begin{equation}
F_{ab}= e_{a}^{\mu}D_{\mu}A_{b}-e_{b}^{\mu}D_{\mu}A_{a},
\label{FISH}
 \end{equation}
where $D_{\mu}A_{a}=\partial_{\mu}A_{a}+\omega_{\mu ab}A^{b}$,
we choose the following matter Lagrangian for massless vector bosons
\begin{equation}
L_{v.b.}=-\frac{1}{4}\eta^{ac}\eta^{bd}F_{ab}F_{cd}
\label{Fab}
  \end{equation}
	In the first order formalism it is understood that $\omega_{\mu bc}$ 
is regarded as an independent variable, to be determined from the 
equations of motion obtained by variating $\omega_{\mu bc}$. Notice 
that the matter Lagrangian (\ref{Fab}) is in fact homogeneous of degree 2 
in the vielbeins. Therefore a theory incorporating only $L_{v.b.}$ as a 
matter model, is consistent with the local Einstein symmetry 
(\ref{LAS1})-(\ref{LAS3}) and satisfies in an automatic 
form, the constraint(\ref{C}) with $M=0$. As a consequence of the 
symmetry (\ref{LAS1})-(\ref{LAS3}), we obtain of course that 
in this model $\chi$ can be taken to be $1$ if $\Phi\neq 0$ everywhere. 
If we do this, then we can see immediately that in the approximation where 
$\omega_{\mu 
ab}=\omega_{\mu ab}(e)$ ($\omega_{\mu ab}(e)$ is the Riemannian 
spin-connection (see also eq.(\ref{RK})), the Lagrangian density (\ref{Fab}) 
with $F_{ab}$ defined by (\ref{FISH}) is invariant under the gauge 
transformations  
\begin{equation}
A_{a}\longrightarrow A_{a}+e_{a}^{\mu}\frac{\partial\Lambda}{\partial 
x^{\mu}}
 \label{GT}
  \end{equation}
If we do not fix $\chi$-field, then the form of the gauge transformations 
is modified, but the model is still gauge invariant in the same 
approximation.

We should point out that together with the obvious advantages which this
formulation of the theory of massless vector bosons has, this approach 
leads to weak violations (of gravitational strength) of the gauge 
invariance principle. This is a consequence of the first order formalism, 
where the 
spin-connection is determined from its equation of motion and we obtain in 
fact  that there will be a contribution to the torsion from the vector 
boson itself. For example, it turns out that this gravitational back 
reaction of the 
vector bosons on gravity (i.e. on $\omega_{\mu ab}$) is propotional to the 
gravitational constant $\kappa$ and violates the gauge invariance. 
Contribution to $\omega_{\mu ab}$ from fermions would also produce 
violations of gauge invariance in (\ref{Fab}).

\bigskip
\subsection{Gauge Fields from Extra Dimensions in the VSC approach}

\bigskip
Here we will see that in the framework of higher dimensional unification, 
the VSC-approach can incorporate gauge fields. It is important to notice 
that in the context of the NGVE-theories only the VSC alternative 
can 
successfully implement the idea of higher dimensional unification. The 
Riemannian approach developed in \cite{GK} and reviewed in section 3, is 
not suitable for this task.

Let us see first of all that the purely Riemannian approach to the 
NGVE-theories does not provide a successful formulation  of the higher 
dimensional unification. To see this, we start from the higher 
dimensional NGVE-Riemannian action  

\begin{equation}
S_{5}=-\frac{1}{\kappa_{5}}\int\Phi R_{5}(\gamma_{ab})d^{5}x,
\label{A5}
  \end{equation}
where
\begin{equation}  
\Phi \equiv 
\varepsilon_{abcde} \varepsilon^{ABCDE}
(\partial_{A}\varphi_{a})
(\partial_{B}\varphi_{b}) \ldots
(\partial_{E}\varphi_{e}),
\label{Fish5}
\end{equation}
$\gamma_{ab}$ is the 5-dimensional metric and $R_{5}$ is the Riemannian 
scalar curvature in the 5-dimensional space-time.

Our choice of parametrizing the 5-dimensional metric $\gamma_{AB}$ 
is\cite{KK}

\begin{displaymath}
\gamma_{AB}=
\left(\begin{array}{ccc}
 g_{\mu\nu}+e^{2}\kappa^{2}v A_{\mu}A_{\nu}&
             e\kappa v A_{\mu}\\
e\kappa v A_{\nu}&
             v\\
\end{array} \right)
\end{displaymath}

where $g_{\mu\nu}$, $v$ and $A_{\mu}$ do not depend on the fifth 
dimension $x^{5}$, which is taken to be compactified. Doing the 
integration over $x^{5}$ we get
\begin{equation}
S_{5}=\frac{1}{\kappa}\int\Phi[-R_{4}+e^{2}v g^{\mu\rho}g^{\nu\sigma}
F_{\mu\nu}F_{\rho\sigma}]d^{4}x,
\label{A54}
  \end{equation}
where $R_{4}$ is the scalar curvature of a 4-dimensional space-time with 
the metric $g_{\mu\nu}$, $\kappa=\kappa_{5}/2\pi\rho$, $\rho$ being the 
size of the extra 
dimension and $F_{\mu\nu} =\partial_{\mu}A_{\nu}-\partial_{\nu}A_{\mu}$.
 
We see now that in contrast with the usual Kaluza-Klein theories, 
variation with respect to $v$ leads to the nontrivial constraint 
for the gauge field $g^{\mu\rho}g^{\nu\sigma}F_{\mu\nu}F_{\rho\sigma}=0$.
Such constraint is of course inconsistent with a phenomenologically 
successful theory of gauge fields, showing therefore the failure of the 
Riemannian approach to the Kaluza Klein unification in the context of the 
non gravitating vacuum energy theories. We now turn our attention to 
higher dimensional unification in the context of the VSC-approach.

Let us consider then the action (\ref{Action}) in the five 
dimensional case ($D=5$), but where the scalar curvature is defined by
(\ref{Scurv}), which means that $R$ may not coincide with the Riemannian
definition, as we have verified it is the case in the four dimensional 
theory. 

Let us now consider the dependence of the spin connection $\omega_{\mu ab}$ 
on the vielbein $e_{a}^{\mu}$ and on 
$\chi\equiv\frac{\Phi}{\sqrt{\gamma}}$ ( we follow the same steps we went 
through in the four dimensional case). Variation of the action 
(\ref{Action}) with $R$ defined as in 
(\ref{Scurv}) in five dimensions gives

\begin{equation}
\varepsilon_{abcdf} \varepsilon^{ABCDF}
(\chi e_{C}^{c}e_{D}^{d}D_{B}e_{F}^{f}
+\frac{1}{3}e_{C}^{c}e_{D}^{d}e_{F}^{f}\chi ,_{B})=0
\label{e5}
\end{equation}

The solution of eq.(\ref{e5}) is now
\begin{equation}
\omega_{A}^{ab}=\omega_{A}^{ab}(e)+K_{A}^{ab},
\label{SC5}
\end{equation}  
where $\omega_{A}^{ab}(e)$ is the Riemannian spin-connection of the 
5-dimensional space-time and 
$K_{A}^{ab}$ is the contorsion tensor which in our case is given by
\begin{equation}
K_{A}^{ab}=\frac{1}{3}\sigma ,_{B}(e_{A}^{a}e^{bB}+e_{A}^{b}e^{aB}),
\label{K5}
\end{equation}

If we insert this into the expression for $\Phi R_{5}(e,\omega)$, we obtain
\begin{equation}
\Phi R_{5}(e,\omega)\equiv\sqrt{\gamma}\chi R_{5}(e,\omega)=
\sqrt{\gamma}[\chi R_{5}(\gamma_{AB})-
\frac{16}{3}\chi^{1/2}\Box\chi^{1/2}]=0.
\label{CI5}
\end{equation} 
Here $R_{5}(\gamma_{AB})$ is the ordinary scalar curvature in the 
5-dimensional Riemannian space-time with the metric $\gamma_{AB}$.
Again, we find a conformal coupling appropriate to $D=5$, for the field 
$\chi^{1/2}$.

The other equations of motion, obtained from the variation with respect 
to $e_{a}^{A}$, after some algebraic manipulations, are

\begin{equation}
G_{(5)AB}(e,\omega)=G_{(5)AB}(\gamma_{CD})+H_{(5)AB}(\chi^{1/2})=0,
\label{EE5}
\end{equation}
where $G_{(5)AB}\equiv R_{(5)AB}-\frac{1}{2}\gamma_{AB}R_{5}$ and

\begin{equation}
H_{(5)AB}(\chi^{1/2})=\frac{2}{\chi^{1/2}}[\gamma_{AB}\Box\chi^{1/2}-
\frac{2}{3}(\chi^{1/2}),_{A;B}]-
\frac{2}{3\chi}[\gamma_{AB}(\chi^{1/2}),_{C}
(\chi^{1/2})^{,C}-
2(\chi^{1/2}),_{A}
(\chi^{1/2}),_{B}]
\label{H5}
 \end{equation}

As in the four dimensional case, if $\Phi\neq 0$, using the symmetry 
(\ref{LAS1})-(\ref{LAS3}) (which in terms of $\chi$ and 
$\gamma_{AB}$ appears as a conformal transformations, see  
(\ref{LSI1})-(\ref{LSI3}), where $g_{\mu\nu}$ should be 
replaced 
by $\gamma_{AB}$), we can set the gauge $\chi=1$, obtaining then 
equations identical to those of ordinary general relativity, in this case 
for $D=5$ however. The Kaluza-Klein mechanism for gauge field generation 
works then as usual.

When considering nonabelian compactifications, things work most 
straightforward when the matter Lagrangian that produces the 
compactification satisfies the constraint (\ref{C}). This is the case if 
the compactification is achieved, for example, through some hedgehog 
configuration\cite{HH} which corresponds to the identity mapping from the 
extra 
dimensional sphere into a space of scalar fields satisfying nonlinear 
sigma model type equations. In addition, instead of sphere some other 
finite area noncompact manifolds can be considered\cite{GZ}. The problem in 
this case is associated with the large mass generation which the 
Kaluza-Klein gauge fields get in these kind of approaches.

\bigskip
\section{Breaking local Einstein symmetry}

\bigskip

\subsection{A Gauge+Matter fields System}

\bigskip
 
In some cases, the constraint (\ref{C}) is not automatically satisfied in 
fact. In those cases, in order for the constraint to be satisfied, the 
$\chi$-field becomes determined, therefore breaking the symmetry 
(\ref{LAS1})-(\ref{LAS3}).

To see how this works, we study a model of a gauge field ( now formulated 
in the GVLS way, in contrast to what we did in section 5.2) and a neutral 
scalar field. Now gauge invariance is evident, however local Einstein 
symmetry is broken. Although the model is not very realistic we study it 
only to get an insight on how the theory works rather than to get a 
correct description of nature.

Therefore we study the model (\ref{Action}) with the particular choice of 
the matter Lagrangian density $L_{m}$ given by 

\begin{equation}
L_{m}=\frac{1}{2}g^{\mu\nu}\partial_{\mu}\phi\partial_{\nu}\phi-
U(\phi)-\frac{1}{4}F_{\mu\nu}F_{\alpha\beta}g^{\mu\alpha}g^{\nu\beta},
\label{LGS}
\end{equation}
where $F_{\mu\nu}\equiv\partial_{\mu}A_{\nu}-\partial_{\nu}A_{\mu}$.

Notice that the action (\ref{Action}) with the matter Lagrangian 
(\ref{LGS}) is {\em not\/} invariant under the local Einstein symmetry 
(\ref{LAS1})-(\ref{LAS3}). However the nontrivial constraint

\begin{equation}
-\frac{1}{2}F_{\mu\nu}F^{\mu\nu}+2[U(\phi)+M]=0
\label{NTC}
\end{equation}

is still satisfied as a result of the equations of motion.

We can study now how the theory works in several types of solutions. 
First of all, if we are interested in radiation type solutions, where 
$F_{\mu\nu}F^{\mu\nu}=0$, the situation becomes identical (from the 
point of view of symmetries) to that  when no gauge field was 
considered (see section 4.4).

If we look for example for static purely electric spherically symmetric 
solutions of the equations of motion, equation (\ref{NTC}) tells us that 
$\phi$ is a function of the electric field, not just a constant as in
the section 4.4. After this, the equation of motion for the $\phi$ -field 
(\ref{SE}) allows us to solve $\frac{d\sigma}{dr}$ as a function of 
$\phi=\phi(F_{0r})$ and its first and second derivatives (as well as 
function of the metric). Finally this solution for $\frac{d\sigma}{dr}$ 
has to be inserted in the equation for $E\equiv F_{0r}$, which involves 
$\frac{d\sigma}{dr}$. The resulting problem is a highly nonlinear one but 
a well defined one which shows the role of the  
$\sigma\equiv\ln\chi$-field in the enforcement of the constraint. Notice 
that the $\chi$-field  is now not arbitrary. However, away from the 
sources, that is in a vacuum state that satisfies constraint (\ref{CS}), 
equation (\ref{SS}) holds, which is an equation with conformal invariance.
Therefore $\chi$ is there totally arbitrary and therefore unphysical.

\bigskip

\subsection{Breaking the local Einstein symmetry by the gravitational 
sector of the theory}

\bigskip

Breaking of the local Einstein symmetry is possible also in the 
gravitational sector of the theory in a case of an appearance of higher
 order 
terms in the curvature (for example as could be the case for quantum 
corrections). These terms usually give rise to non causal 
propagation and ghosts. In our case however, the fact that the measure of 
integration is $\Phi$ instead of $\sqrt{-g}$ allows us to consider 
contributions which are meaningless in the usual theory. This is the case 
when in the Lagrangian density we consider possible Euler ($\rho$) and 
Hirzebruch-Pontryagin ($\xi$) contributions 

\begin{equation}
\rho\equiv\frac{1}{\sqrt{-g}} 
\varepsilon^{\alpha\beta\mu\nu}\varepsilon^{abcd}
R_{\alpha\beta ab}R_{\mu\nu cd}
\label{Eul}
\end{equation}

\begin{equation}
\xi\equiv\frac{1}{\sqrt{-g}}\varepsilon^{\alpha\beta\mu\nu}
R_{\alpha\beta ab}R_{\mu\nu cd}\eta^{ac}\eta^{bd}
\label{Pont}
\end{equation}

$\rho\sqrt{-g}$ and $\xi\sqrt{-g}$ are total divergences even 
in the presence of torsion\cite{Ni} (our conventions for 
$\varepsilon^{\alpha\beta\mu\nu}$ are different from those of
 Ref.\cite{Ni})
 and therefore irrelevant in the 
standard approaches. However, since the measure of integration is $\Phi$,
$\rho\Phi$ and $\xi\Phi$ are  contributions to the action 
that can be considered in our case. These give nontrivial 
contributions to the equations of motion. Since the contribution of $\xi$ 
into the total action violates the 
parity symmetry, we do not consider it in this first analysis. Furthermore,
 we present here only a sketch about the main features of the 
theory in the presence of the Euler contribution into the Lagrangian 
density.

It is known\cite{Z} that when studying space-time of dimensionalities 
bigger than four, the corresponding generalization of the 
four-dimensional Euler density gives  a nontrivial contribution to
the equations of motion, however does not give rise to ghosts.
In our case, this is still true and the proof follows the same lines 
of what is done in the higher dimensional case\cite{Z}.

Considering small perturbations of $e_{a}^{\mu}$ and $\chi$ around 
$e_{a}^{\mu}=\delta_{a}^{\mu}$ and of $\chi =1$, we obtain from the Euler 
contribution into the Lagrangian density:

\begin{equation}
S_{E}=\int \Phi d^{4}x\partial_{\alpha}j^{\alpha},
\label{EAL}
\end{equation}
where 

\begin{equation}	
j^{\alpha}=4[2\sigma^{,\alpha}_{;\beta}\sigma^{,\beta}-
2\sigma^{,\alpha}\Box\sigma-
\sigma^{,\beta}\sigma,_{\beta}\sigma,_{\alpha}]
\label{ECar}
\end{equation}
and $\sigma =\ln\chi$.
As in Ref.\cite{Z}, the purely gravitational effects vanish in the 
quadratic approximation. We see, for example, that
$(\frac{\partial^{2}\sigma}{\partial t^{2}})^{2}$ is absent in the 
integrand of eq. 
(\ref{EAL}).
Second derivatives with respect to time appear in the integrand 
of eq. (\ref{EAL}) 
only linearly.

Notice that when the Euler density is present, the constraint (\ref{C})
becomes now a dynamical equation for $\sigma$:

\begin{equation}
2\rho+e^{a\mu}\frac{\partial L_{m}}{\partial e^{a\mu}}-2L_{m}+2M = 0,
 \label{CEE}
 \end{equation}

where $\rho$ is given by eq.(\ref{Eul}). Eq. (\ref{CEE}) is a dynamical
equation for $\sigma$ rather than a constraint because second order time 
derivatives of $\sigma$ appear in (\ref{CEE}).

Finally notice that in the context of the modifications in the 
gravitational sector, described in this subsection,
 flat space-time is still 
always a solution.

.\bigskip
\subsection{General Relativity limit as the freezing the $\chi$ degree of 
freedom}

\bigskip

As we have seen, the vanishing of the cosmological constant relies on the 
existence of the $\chi$-field and the nontrivial constraint (\ref{C}) 
associated 
with the new measure of integration and with the scalar fields 
$\varphi_{a}$ from which this measure (and the $\chi$-field) is build.

Now we want to see in what limit the model studied here becomes 
undistinguishable from GR. This is important from the point of view of 
the correspondence principle. This question has obviously to do with what 
is assumed for the dynamics of the $\chi$-field.

When the local symmetry (\ref{LAS1})-(\ref{LAS3}) exists, the 
$\chi$-field does not represent a physical degree of freedom and it is in 
fact arbitrary and not determined by the equations of motion. When the 
constraint (\ref{C}) is nontrivially satisfied, the $\chi$-field has 
to be determined so as to make things work. Finally, as we explained in 
section 6.2, we found a way to turn the $\chi$-field into a dynamical 
field by introducing the Euler term.

Obviously we should be able to obtain GR if the dynamics of the 
$\chi$-field is turned into a trivial one, that is, if only $\chi 
=constant$ is allowed. This is of course the exact opposite to the case 
of unbroken symmetry. This is possible within the general form 
(\ref{Action}) of the theory, provided we add to $\L_{m}$ a Lagrange 
multiplier term that enforces $\chi =constant$. The form of this 
contribution to the Lagrangian density is      

\begin{equation}
\L_{freezing}=\frac{1}{\sqrt{-g}}\varepsilon^{\mu\nu\alpha\beta}
\partial_{\mu}E_{\nu\alpha\beta}
\label{Freez}
\end{equation}

where we assume that all components of $E_{\nu\alpha\beta}$ are to be varied 
without restriction, i.e. $E_{\nu\alpha\beta}$ is a new fundamental field.
The variation of the action with respect to $E_{\nu\alpha\beta}$ gives in 
fact
\begin{equation}
\partial_{\mu}\chi = constant,
\label{constantchi}
\end{equation}
that is the only possible configuration for the $\chi$-field is  $\chi 
=constant$.

 The variation of the action with respect to $e^{a\mu}$ gives

\begin{equation}
 -\frac{2}{\kappa}R_{a\mu}(\omega)
+\frac{\partial L_{m}}{\partial e^{a\mu}} +e_{a\mu} L_{freezing} = 0,
 \label{ETfreeze}
 \end{equation}

Contracting this with $e^{a\mu}$ we obtain

\begin{equation}
 -\frac{2}{\kappa}R(\omega)
+e^{a\mu}\frac{\partial L_{m}}{\partial e^{a\mu}} +4 L_{freezing} = 0
 \label{COLD}
 \end{equation}

Also, the variation with respect to the fields $\varphi_{a}$
leads to ( if $\Phi \neq 0$) 

\begin{equation}
 -\frac{1}{\kappa}R(e,\omega ) + L_{m}(e,\omega, matter fields)+ 
L_{freezing}=M 
 \label{Minibus}
\end{equation}
From (\ref{COLD}) and (\ref{Minibus}) we get
\begin{equation}
e^{a\mu}\frac{\partial L_{m}}{\partial e^{a\mu}}-2(L_{m} - 
L_{freezing}-M) = 0
 \label{FrC}
 \end{equation}

which now is not a constraint but rather determines the Lagrange multiplier
term $L_{freezing}$.

Contracting (\ref{ETfreeze}) with $e^{a}_{\nu}$ we obtain

\begin{equation}
 -\frac{2}{\kappa}R_{\mu\nu}(\omega)
+e^{a}_{\nu}\frac{\partial L_{m}}{\partial e^{a\mu}} 
+g_{\mu\nu}L_{freezing} = 0
 \label{Rfr}
 \end{equation}

From (\ref{Rfr}) and (\ref{COLD}) we get 
\begin{equation}
 -\frac{2}{\kappa}[R_{\mu\nu}(\omega)-\frac{1}{2}g_{\mu\nu}R(\omega)]
+e^{a}_{\nu}\frac{\partial L_{m}}{\partial e^{a\mu}}
-g_{\mu\nu} L_{m}+g_{\mu\nu}M= 0,
 \label{EnFr1}
 \end{equation}

or what is the same
\begin{equation}
 G_{\mu\nu}=\frac{\kappa}{2}
(e^{a}_{\nu}\frac{\partial L_{m}}{\partial e^{a\mu}}
-g_{\mu\nu} L_{m}+g_{\mu\nu}M)= 
\frac{\kappa}{2}(T_{\mu\nu}+Mg_{\mu\nu}),
 \label{EnFr2}
 \end{equation}
where $G_{\mu\nu}\equiv R_{\mu\nu}-\frac{1}{2}g_{\mu\nu}R$.

Taking into account that the $\chi$ degree of freedom is frozen now(see 
eq.(\ref{constantchi})) we conclude that  eqs.(\ref{EnFr2}) are the Einstein
equations of GR (or Einstein-Cartan equations if fermions are included in 
the model), where an arbitrary integration constant $M$ playing the role 
a cosmological constant appears.

 No information concerning the vanishing 
of the cosmological constant is obtained now, because the constraint that 
used to do this job contains now the Lagrange multiplier field 
$E_{\nu\alpha\beta}$ which is not determined by any other equation, 
therefore eq.(\ref{FrC}) is not now a constraint at all.

\bigskip
\section{Discussion and Conclusions}

\bigskip

Here we have developed the consequences of changing the measure of 
integration from $\sqrt{-g}d^{D}x$ to
 $d\varphi_{1}\wedge d\varphi_{2}\wedge\ldots\wedge d\varphi_{D}$,
 when the 
mapping from the scalars $\varphi_{a}$ ($a=1,2,\ldots D$) to 
the coordinates is not known a priori. This means that the measure of 
integration is determined dynamically and not assumed to have a 
particular form as it is done in GR. Such model\cite{GK} has been called
"Nongravitating vacuum energy (NGVE) theory" because if we change the 
integrand (i.e. the Lagrange density) by a constant, which in GR is 
associated with a vacuum energy, no change in the equations of motion is 
obtained (ignoring possible boundary effects).

Moreover, we have discovered in this 
paper, that when using the vielbein - spin-connection formalism in the 
context of the models based on the NGVE-principle, a nontrivial 
constraint (\ref{C}) appears as a result of equations of motion.

In our previous paper on NGVE-theory\cite{GK} we have seen that the 
constant part of the vacuum energy does not affect gravitational 
properties, but an integration constant appears and it plays a role 
similar to that of an effective cosmological term (although a maximally 
symmetric de Sitter space did not exist there). Now, allowing for the 
possibility of torsion, such integration constant appears too, but it is 
determined {\em dynamically\/} so as to cancel any possible constant part 
of the vacuum energy, which is present in the starting formulation.

The existence of a nontrivial constraint does however modify the 
dynamics of the matter fields in a nontrivial way. This can be avoided, 
as we have seen in section 6.4, by making the dynamics of the $\chi$-field 
trivial if we introduce a Lagrange multiplier. In this case the 
constraint becomes only a definition of the Lagrange multiplier and 
therefore fails to give additional information on the effective 
cosmological constant. This version of the theory coincides physically 
with the models discussed by the authors of Refs\cite{Sqrtg} where the 
cosmological constant is an integration constant. It is here obtained 
from the general formalism by enforcing from outside the triviality of 
the $\chi$-field, which is of course not a natural things to do.
This version of the theory does not answer the question of the vanishing 
of the cosmological constant, which is again an arbitrary integration 
constant, with no particular reason to pick the vanishing value as has 
been pointed out by Weinberg\cite{WNg}.

We have seen in section 6.1 that terms that violate the local Einstein
symmetry can be incorporated and they do not alter the basic conclusions
of the model, that is, the vanishing of the cosmological constant term.
They do give rise however to a nontrivial dynamics for the field $\chi$
which acquires a physical meaning due to these breaking terms. Furthermore,
the constraint (\ref{C}) is satisfied anyway by dynamically adjusting the
field $\chi$, as we saw in the particular example of section 6.1.

Incorporating masses for example, will modify this constraint so that the
masses will enter in the constraint. In the case of fermions, we have
seen in section 4.3
that if we start from  Nambu-Jona Lasinio (NYL) type models\cite{NJL},
the constraint(\ref{C}) is satisfied on the mass shell without
restriction on
the matter field dynamics. However a
spontaneous symmetry breaking mechanism originates from
quantum corrections and as a result masses of fermions appear. So if our
classical arguments
concerning the satisfaction of the constraint(\ref{C}) in the NJL model
survive the
quantum corrections, we would then expect that the fermion masses may not
enter in the constraint at all. If they do, they contribute to a non
trivial $\chi$ dynamics. Which alternative is the right one requires a
nontrivial analysis.

In addition , in the context of some model resembling the standard model,
the constraint(\ref{C}) seems to give a basic condition which tell us that
the
Higgs field is a composite of the other fields appearing in the theory in
a way that resembles what we have studied in section 6.1.

A way to avoid the constraint from having a big effect 
on the dynamics of any single matter field is to introduce a large number 
of fields, most of them interacting with each other only gravitationally 
and of course through the constraint. Since the whole $L_{m}$ enters in 
the constraint, enlarging the number of fields diminishes the "job" each 
individual field has to do. In such a way  we expect to recover the local 
symmetry (\ref{LAS1})-(\ref{LAS3}) at long distances, i.e. the triviality of 
the 
constraint at long distances. This would be a way to realize the infrared 
dynamical symmetry restoration of gauge symmetries as it has been 
discussed in the literature\cite{Nl}.
 
Keeping a nontrivial constraint (that is avoiding the introduction of the 
Lagrange multiplier that trivializes the $\chi$-dynamics), in sections 
3,4 and 5 we have formulated several models (including fermions, scalar 
field and vector bosons) where the constraint (\ref{C}) is satisfied 
at least on the mass shell. However, it follows from the equations 
of motion that the only possible configuration of a single scalar field 
with a 
potential is a constant scalar field located in the extremum of the 
potential. In this case {\em the constraint dictates that this extremal 
value of the potential is compensated by the integration constant, thus 
providing the mechanism for the nonexistence of the cosmological constant 
on the mass shell\/}.  Those models respect the local Einstein 
symmetry. Therefore we can set the gauge $\chi =1$ and in this case
 eq.(\ref{Nonconse}) becomes the equation of the covariant conservation 
of the energy-momentum tensor. 

The infinite dimensional symmetries (\ref{infphi}) and (\ref{L}) impose 
strict restrictions on the possible induced terms in the quantum 
effective action, if no anomalies appear in this effective action. In 
particular, symmetry under the transformations (\ref{L}) seems to prevent 
the appearance of terms of the form $f(\chi)\Phi$ (except for
 $f(\chi)\propto 1/\chi$) in the effective action which although is 
invariant under volume preserving transformations (\ref{infphi}), breaks 
symmetry (\ref{L}). The case $f(\chi)\propto 1/\chi$ is not
 forbidden by symmetry (\ref{L}) and appearance of such a term would mean
inducing a "real" cosmological term, i.e. a term of the form 
$\sqrt{-g}\Lambda$ in the effective action. However, appearance of such a 
term seems to be ruled out because of having opposite parity properties 
to that of the action given in (\ref{Action}). Furthermore, in the absence 
of Euler-like terms (of section 6.2), the variational principle gives 
now  $\Lambda =0$ in the vacuum if such term is "forced" into the theory. 
Of course, in the absence of a consistently quantized theory, such 
arguments are only preliminary. Nevertheless it is interesting to note 
that if all these symmetry arguments are indeed applicable, this would
 imply that the scalar fields $\varphi_{a}$ can appear in the effective 
action only in the integration measure, that is they preserve their 
geometrical role.

Finally, it is very interesting that in attempts to build a model which 
respects both the local Einstein symmetry and the gauge invariance, we 
have succeeded in finding it only in the framework of the Kaluza-Klein 
unification. It is a clue that the resolution of the cosmological 
constant problem and the problem of unifying the fundamental forces of 
nature are intrinsically intertwined.    

\bigskip

\section{ Acknowledgements}

We would like to thank J.Bekenstein, R.Brustein, A.Davidson, D.Owen, 
L.Parker and Y.Peleg for interesting conversations. Specially we want to 
thank L.Parker and Y.Peleg for interesting suggestions concerning the 
special role of Kaluza-Klein unification in the context of our model.

\pagebreak

\end{document}